# Pushing Configuration-Interaction to the Limit: Towards Massively Parallel MCSCF Calculations

Konstantinos D. Vogiatzis[1*,†], Dongxia Ma[1¶], Jeppe Olsen[2], Laura Gagliardi[1]*, Wibe de Jong[3]*

[1] *Department of Chemistry, Minnesota Supercomputing Institute, and Chemical Theory Center, University of Minnesota, 207 Pleasant Street Southeast, Minneapolis, MN 55455-0431, United States*

[2] *Department of Chemistry, Aarhus University, Langelandsgade 140, 8000 Aarhus C, Denmark*

[3] *Computational Research Division, Lawrence Berkeley National Laboratory, Berkeley, California 94720, United States*

**Abstract**

A new large-scale parallel multiconfigurational self-consistent field (MCSCF) implementation in the open-source NWChem computational chemistry code is presented. The generalized active space (GAS) approach is used to partition large configuration interaction (CI) vectors and generate a sufficient number of batches that can be distributed to the available nodes. Massively parallel CI calculations with large active spaces can be treated. The performance of the new parallel MCSCF implementation is presented for the chromium trimer and for an active space of 20 electrons in 20 orbitals. Unprecedented CI calculations with an active space of 22 electrons in 22 orbitals for the pentacene systems were performed and a single CI iteration calculation with an active space of 24 electrons in 24 orbitals for the chromium tetramer was possible. The chromium tetramer corresponds to a CI expansion of one trillion SDs (914 058 513 424) and is largest conventional CI calculation attempted up to date.

## 1. Introduction

The accurate calculation of near-degeneracy electron correlation effects for large orbital spaces is central in modern electronic structure theory. Many systems of interest cannot be quantitatively described by a single electronic configuration. Multireference effects, also referred as static correlation, nondynamic correlation, left-right correlation or strong correlation,[1-3] can be captured by the full configuration interaction (full CI) expansion of the wave function. In full CI theory, the wave function is a linear expansion of all the Slater Determinants (SDs) or spin-adapted configuration state functions (CSFs) that can be generated in a given one-electron basis. The exponential dependence of the number of SDs on the number of orbitals and electrons makes full CI wave function applicable to only small to modest sized systems.

In multiconfigurational self-consistent field (MCSCF) theories a full CI is employed on a selected orbital subspace (*active* orbitals), while the remaining orbitals are kept either occupied (*inactive*) or empty (*virtual* or *secondary*), and orbitals are variationally optimized simultaneously with the configuration expansion coefficients.[4] These two problems are usually decoupled and solved separately. In the inner loop (microiterations) the CI coefficients are optimized minimizing the energy. In outer loops (macroiterations) the molecular orbitals are optimized by an iteratively solution of the Newton-Raphson equations using the first-order density matrix calculated from the CI expansion. A full CI calculation in the active space is performed at every MCSCF iteration and thus, considerable effort has been performed over the past 40 years to develop and implement efficient CI algorithms.[4-8]

The CI-related methods have taken advantage of parallel architectures, and significant progress has been made in the last 20 years.[4, 9-16] However, new architectures with increased parallelism require algorithmic improvements of parallel CI and MCSCF implementations to take full advantage of these technological advances. Current parallel CI calculations are able to tackle expansions of a few billions of CSFs.[4] To the best of our knowledge, the largest multireference CI (MRCI) calculation that has been reported is 2.8 billion CSFs (60 billions SDs)[17], while the largest CI expansion in a full CI calculation contains 10 billion determinants.[9]

Alternative expansions of the wave function have been proposed that allow access to larger active spaces while limiting the number of determinants in the CI expansion. The density-matrix renormalization group (DMRG)[18-20] substitutes the exact diagonalization of large Hamiltonian matrices by encoding a sequential structure into the correlation. The DMRG wave function is built from local variational objects associated with the active orbitals of the system. The DMRG-SCF methodology allows the effective treatment of large molecular complexes and is gradually becoming a standard quantum chemical method.[21-22] The variational two-electron reduced-density matrix (v2RDM) method and the corresponding v2RDM-CI and MCSCF variants have been recently applied for solving strongly correlated systems.[23-25] Stochastic approaches have been suggested as an efficient alternative to the standard Davidson CI eigensolver.[26-27] The full CI quantum Monte Carlo (FCIQMC)-MCSCF method has been applied to study transition metal complexes such as Fe porphyrins.[28-29]

The restricted active space SCF (RASSCF)[7], the generalized active space SCF (GASSCF)[30], and the occupation restricted multiple active space (ORMAS)[31] methods provide a different approach to the reduction of the CI expansion by limiting the excitations within the active orbitals.[7, 30, 32] In the GASSCF approach, multiple orbital spaces are chosen instead of one complete active space (CAS). The definition of the intra- and interspace electron excitations leads to an efficient elimination of negligible configurations from the configuration space, which effectively reduces the CI expansion. The spawning of multiple active spaces provides an approach to split the CI vector into smaller blocks (*vide infra*).[7, 33-34]

The most time-consuming step of a CI calculation is the construction of the σ vector needed in the Davidson algorithm. This step is also the most difficult part in terms of parallelization, as it is subject to load-balance, bandwidth and memory constrains. In this study we used the GAS framework for the development of a new implementation of a massively parallel MCSCF code. The new parallel MCSCF implementation, based on a serial version of the LUCIA[34] program, was efficiently parallelized and integrated into the NWChem program package.[35] This implementation allows us to perform large-scale CI calculations with a fast time-to-solution, as well as allow us to explore active spaces beyond the limits of conventional MCSCF implementations.

The outline of the paper is as follows: In section 2, the foundations of MCSCF theory are discussed. In section 3, the technical aspects of the parallel MCSCF implementation are presented. The performance of this implementation for an active space with 20 electrons in 20 orbitals is presented in section 4. In section 5 the applicability of the new parallel code to larger full CI spaces, and possible further improvements to the parallel performance are discussed. Finally, in section 6 we offer some conclusions

**2. Theory and Methodology**

In CASSCF theory, the size of the CI expansion is dictated by the size of the complete active space, or CAS. The choice of the number of orbitals and electrons that compose the active space is usually system dependent and is based on the nature of the chemical problem under consideration. The number of SDs included in the CI expansion scales exponentially with the size of the active space and active spaces larger than 18 electrons in 18 orbitals cannot be currently treated.[36]

Restricting excitations between orbitals in the generation of the CI expansion lead to the reduction of the number of SDs or CSFs that need to be considered. Such restrictions can usually be rationalized by the chemistry of the molecular system, and are system dependent, but can lead to a simplification of the CI problem without significant loss of accuracy.

2.1. Determinant-based Direct-CI

In full-CI theory, and for a given one-electron expansion, the exact solution of the Schrödinger equation may be written as a linear combination of all Slater determinants that can be constructed in the $N$-electron

Fock space. The MCSCF wave function in which the SD basis is constructed from a subspace (active space) of the full Fock space is expressed as:

$$|\Psi_{\text{MCSCF}}\rangle = \exp(-\hat{\kappa}) \sum_i C_i |\Psi_i\rangle \quad (1)$$

where $i$ is the total number of SDs, $C_i$ the variational CI coefficients, and $\exp(-\hat{\kappa})$ the orbital-rotation operator. The CI eigenvalue problem can be solved with the direct-CI approach[37], in which the expansion coefficients are computed in operator form *directly* from the one- and two-electron integrals within an iterative scheme. The handling of the SDs is simplified if each SD is represented as a product of an alpha and a beta string[38]

$$|\Psi_i\rangle = |\alpha(I_\alpha)\beta(I_\beta)\rangle = \hat{\alpha}(I_\alpha)\hat{\beta}(I_\beta)|\text{vac}\rangle, \quad (2)$$

where $\hat{\alpha}(I_\alpha)$ and $\hat{\beta}(I_\beta)$ are ordered products of alpha and beta creation operators, respectively, and $|\text{vac}\rangle$ is the vacuum state. The MCSCF wave function (or CI expansion) can be written as

$$|\Psi_{\text{MCSCF}}\rangle = \sum_{I_\alpha, I_\beta} C(I_\alpha, I_\beta) |\alpha(I_\alpha)\beta(I_\beta)\rangle \quad (4)$$

We are using the *modified inverted-Davidson* algorithm of LUCIA.[7, 34] The Davidson eigensolver[39] iteratively diagonalizes a subspace instead of the full Hamiltonian matrix.

In a direct CI iteration[40], the main computational cost is the construction of the sigma vector

$$\sigma(I_\alpha, I_\beta) = \sum_{J_\alpha, J_\beta} \langle \beta(J_\beta)\alpha(J_\beta)|\hat{H}|\alpha(I_\alpha)\beta(I_\beta)\rangle C(J_\alpha, J_\beta) \quad (4)$$

or, in a matrix notation,

$$\boldsymbol{\sigma} = \boldsymbol{HC}. \quad (5)$$

The non-relativistic Hamiltonian of Eq. (4) and (5) is expressed as

$$\hat{H} = \sum_{kl} h_{kl}\hat{E}_{kl} + \frac{1}{2}\sum_{ij,kl}(ij|kl)(\hat{E}_{ij}\hat{E}_{kl} - \delta_{jk}\hat{E}_{il}) \tag{6}$$

where $\hat{E}_{kl}$ is the one-electron excitation operator

$$\hat{E}_{kl} = a_{k\alpha}^{\dagger}a_{l\alpha} + a_{k\beta}^{\dagger}a_{l\beta}. \tag{7}$$

By inserting the Hamiltonian of Eq. 6 in Eq. 5, we can rewrite the sigma vector as a sum of three terms:

$$\sigma(I_\alpha, I_\beta) = \sigma_1(I_\alpha, I_\beta) + \sigma_2(I_\alpha, I_\beta) + \sigma_3(I_\alpha, I_\beta), \tag{8}$$

where $\sigma_1$ is a column vector with only beta-beta contributions ($I_\alpha = J_\alpha$), $\sigma_2$ is a column vector with only alpha-alpha contributions ($I_\beta = J_\beta$), and $\sigma_3$ includes the alpha/beta couplings. For more details on the form of the three sigma vector terms, and on the efficiency that this splitting introduces, see Ref. [7].

2.2. Orbital Optimization

A second-order Newton-Raphson iterative procedure is applied for the orbital optimization step of the MCSCF macro-iteration, as implemented in LUCIA, and the variational orbital parameters of the vector $\kappa$ are calculated according to (Eq. 1). The energy can be expressed as

$$E(\kappa) = E(0) + \kappa g + \frac{1}{2}\kappa^2 H, \tag{9}$$

where **g** and **H** are the orbital gradient and orbital Hessian, respectively. The stationary points are obtained as solutions to the equation: $\partial E/\partial p_i = 0$. Orbital rotations between the inactive-active and active-virtual orbital spaces are allowed. The full orbital-orbital Hessian is used, without any approximations.

3. Implementation

The Global Arrays Toolkit[41] was used to facilitate the parallelization of LUCIA. The Global Arrays set of tools was co-developed with NWChem as a shared-memory programming interface for distributed data algorithms relevant to the field of computational chemistry. They allow ease of programming and lack of synchronization between processors, considering that the nonlocal data take more time to access, and offer support for both task and data parallelism. The potentially very large CI and sigma vectors, and smaller Fock matrices utilized in the LUCIA CI and MCSCF code are stored in global arrays that are distributed over the memory available on the allocated processors. A maximum of three vectors are stored in memory, whereas additional vectors needed in the Davidson iterative scheme are stored using parallel IO (ParIO) within the native framework in the Global Arrays Toolkit. LUCIA's local data memory registration routine was modified to utilize NWChem's memory allocation process.

An MCSCF calculation can be divided in four tasks: (1) the generation of the CI expansion, (2) the partial atomic orbital (AO) – molecular orbital (MO) integral transformation, (3) the CI eigenvalue problem, and (4) the solution of the second-order Newton-Raphson equations for the orbital optimization step. For small CI expansions, the AO-MO integral transformation is the most CPU-time demanding step. For large CI expansions, the CI eigensolver (usually the Davidson algorithm via the direct-CI method[40]) dominates the computational time of the MCSCF calculation. Prior to this work, the limitation for the selected active space was 18 electrons in 18 orbitals for a singlet ($S = 0$) spin state. The implementation details about the parallelization of each of the four tasks are discussed in the following sections.

3.1 Generation of CI Expansion for Parallel Processing

The main strategy in the parallelization of the CI algorithm, to enable MCSCF calculations with large active spaces, is to distribute CI and sigma vectors into batches. The batches can be subsequently assigned as parallel tasks to different processors to obtain a good load balance of work among the processors. With each processor only having to store a subsection of the CI and sigma vectors, the memory footprint is significantly reduced. For simplicity NWChem's LUCIA version will store the CI and sigma vectors as SDs instead of the CSFs used by other codes.

LUCIA organizes the alpha and beta strings of Eq. 2 into blocks of SDs with the same occupation type (T) and spin symmetry (S). The occupation type is defined according to the distribution of electrons in each GAS space. Spatial and spin symmetry are defined according to the occupation of the spin orbitals in each string. Therefore, each alpha string has a specific TS value:

$$\hat{\alpha}(I_\alpha) = \prod_{i=1}^{N} \hat{\alpha}(I_{i\alpha}) = \prod_{i=1}^{N} \hat{\alpha}(I_{i\alpha}^{T_i S_i}) = \hat{\alpha}(I_\alpha^{T_1 T_2 \cdots T_N S_1 S_2 \cdots S_N}) = \hat{\alpha}(I_\alpha^{TS}) \qquad (9)$$

where $N$ is the number of GAS spaces. Similar expression holds for a beta string. Combinations of alpha and beta TS strings generate SDs with a specific TTSS definition. SDs with same TTSS definition are grouped into TTSS blocks. The TTSS blocks are furthered grouped in TTSS batches.

By default, LUCIA generates only a limited number of batches, which restricts the number of processors among which the workload can be distributed. The approach used here to increase the number of TTSS blocks and batches, and subsequently the number of tasks available for parallel processing, is to fragment a parent (complete or general) active space and distribute the orbitals over additional GAS spaces. The GAS partitioning approach within LUCIA offers an inherent block distribution of both the CI vector and the Hamiltonian to increase the number of tasks that can be distributed over the available processors. The technique is also used in serial calculations, to reduce the batch size and memory footprint.[34] The approach is schematically depicted in Figure 1.

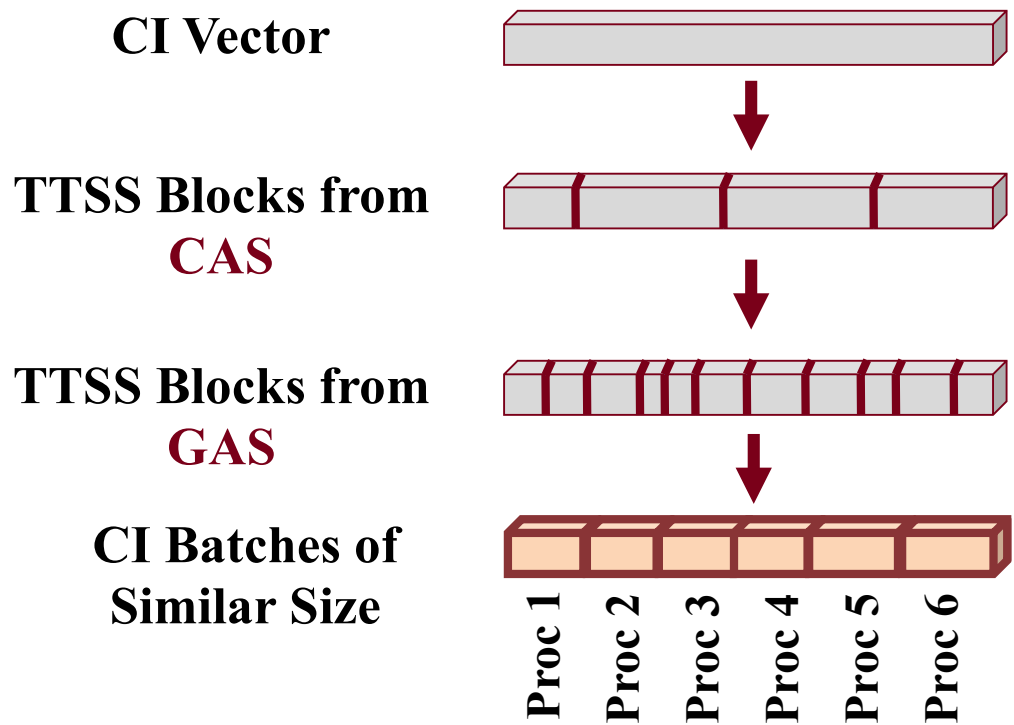

**Figure 1:** Schematic representation of the distribution of the CI vector. Fragmentation of the CI vector (upper line) into a limited number of TTSS blocks for a given CAS (second line). Partitioning of the parent CAS into multiple GAS spaces generates more TTSS blocks (third line), which can be grouped into TTSS batches. The TTSS batches are distributed into the available processors (bottom line).

Dividing the parent active space across multiple GAS spaces can be accomplished in many ways. In the current implementation two strategies have been automated, which are represented in Table 2. The first strategy is to distribute the MOs of each irreducible representation into a separate GAS space (Table 2A). This will increase the number of TTSS batches, but still has limitations, for example when limited point group symmetry is available. The second strategy is an iterative scheme where one MO is moved into a new GAS space (Table 2B). We elected to move an MO from the irreducible representation with the largest number of orbitals. This process is repeated until the number of TTSS batches is larger than the number of processors and a distribution of the CI tasks is feasible. This approach has the additional advantage that it generates smaller TTSS batches that can easily fit in local memory, something that will be crucial for the execution of CI and MCSCF calculations with more than 60 billion SDs. It should be noted that these strategies still have little control over the size and computational intensity of the batches that are created, which will affect our ability to effectively load-balance the work over the processors.

**Table 1:** Representative examples of possible approaches to CAS(20,20) distribution over multiple GAS spaces. (A) Each GAS holds all orbitals from one or more irreducible representations, generating 888 TTSS batches. (B) Individual orbitals from irreducible representations are stored in different GAS spaces, generating 14502 TTSS batches.

| (A) | Irreducible Representation | | | | | | | | (B) | Irreducible Representation | | | | | | | |
|---|---|---|---|---|---|---|---|---|---|---|---|---|---|---|---|---|---|
| | $a_g$ | $b_{3u}$ | $b_{2u}$ | $b_{1g}$ | $b_{1u}$ | $b_{2g}$ | $b_{3g}$ | $a_u$ | | $a_g$ | $b_{3u}$ | $b_{2u}$ | $b_{1g}$ | $b_{1u}$ | $b_{2g}$ | $b_{3g}$ | $a_u$ |
| GAS 1 | 6 | 0 | 0 | 0 | 0 | 0 | 0 | 0 | GAS 1 | 3 | 1 | 1 | 2 | 2 | 2 | 2 | 1 |
| GAS 2 | 0 | 1 | 1 | 2 | 0 | 0 | 0 | 0 | GAS 2 | 1 | 0 | 0 | 0 | 0 | 0 | 0 | 0 |
| GAS 3 | 0 | 0 | 0 | 0 | 5 | 0 | 0 | 0 | GAS 3 | 0 | 0 | 0 | 0 | 1 | 0 | 0 | 0 |
| GAS 4 | 0 | 0 | 0 | 0 | 0 | 2 | 0 | 0 | GAS 4 | 1 | 0 | 0 | 0 | 0 | 0 | 0 | 0 |
| GAS 5 | 0 | 0 | 0 | 0 | 0 | 0 | 2 | 0 | GAS 5 | 0 | 0 | 0 | 0 | 1 | 0 | 0 | 0 |
| GAS 6 | 0 | 0 | 0 | 0 | 0 | 0 | 0 | 1 | GAS 6 | 1 | 0 | 0 | 0 | 0 | 0 | 0 | 0 |
| | | | | | | | | | GAS 7 | 0 | 0 | 0 | 0 | 1 | 0 | 0 | 0 |

Having divided the CI expansion into a large number of TTSS batches, the next step is to distribute the batches over the allocated processors and ensure that the computational work of each processor is balanced and maximum parallelization is achieved. In the design of the parallel algorithm the choice was made to allow access to the data of the sigma batches locally, while CI batches needed in the calculation were fetched using one-sided get operations. All the computational work associated with a TTSS batch is therefore also assigned to the processor where the data resides. This approach significantly reduces the overall communication volume needed, but does require the computed batches to be statically distributed at the beginning of the CI calculation. An alternative approach would be the use of a global task pool, requiring communication of both CI and sigma data blocks if non-locality is assumed.

The order of the TTSS batches in the CI vector is fixed, and it is conceivable that batches have drastically different computational time requirements (see for example the top-left graph in Figure 3), and that computationally expensive batches are located at the beginning or end of the CI vector. One would prefer the number of parallel tasks to be many orders of magnitude larger than the number of processors to balance out the irregular batch sizes and associated computational work. The large number of GAS spaces needed to create, for example, a batch to processor ratio of 100:1, themselves generate a significant computational overhead, as will be shown in section 4. Hence, the challenge in achieving parallel efficiency is finding the optimum balance between the number of GAS spaces and parallel tasks/batches needed given the number of processors.

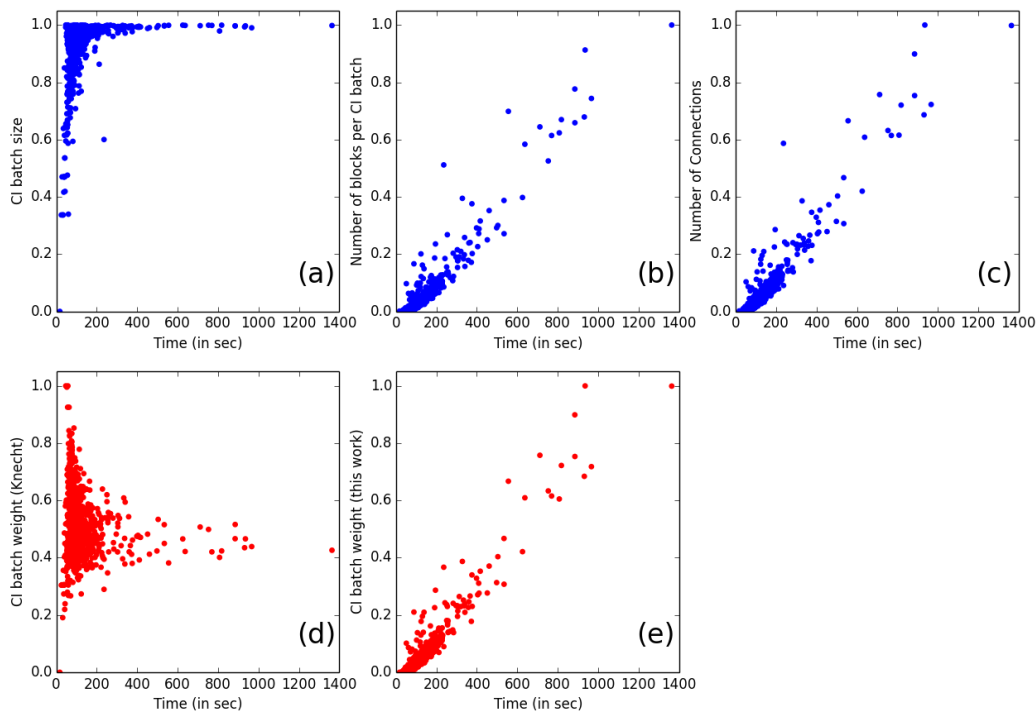

**Figure 2:** Upper line: Scatter plots between the timings of the sigma tasks for each batch and three parameters of the CI batches: (a) batch size in SDs, (b) number of blocks, (c) number of connections. Lower line: scatter plot between two different TTSS weight factors: (d) Knecht et al[14], (e) this work. All y axis values are normalized to 1. All data obtained from a CI expansion of an active space that includes 20 electrons in 20 orbitals (2 133 595 282 SDs, 8 irreducible representations, singlet state).

The work associated with each TTSS batch needs to be estimated accurately and enough batches, i.e. parallel tasks, need to be available to ensure a balanced workload. Three intrinsic parameters of the TTSS (or CI) batches were initially considered to estimate the computational work, and they are plotted versus the timings of the corresponding sigma calculation tasks in Figure 2. These three parameters are the number of SDs or size of the TTSS batch (Figure 2a), the number of TTSS blocks per batch (Figure 2b), and the number of connections between the CI vector and the Hamiltonian (Figure 2c). The connectivity between the CI vector and the Hamiltonian is defined by the electron difference between a TTSS block (subgroup of a TTSS batch) and the Hamiltonian. The diagonal elements differ by zero electrons, while the off-diagonal elements can differ by one or more electrons. The cases that differ by more than two electrons do not couple, and they are not included in the connectivity calculation. All data of Figure 2 were obtained from a CI expansion that includes 20 electrons in 20 orbitals, or CAS(20,20), in a short-hand notation. The size of the CI vector is 2 133 595 282 SDs (linear $Cr_3$ molecule, $D_{2h}$ symmetry with 8 irreducible representations, singlet $A_g$ state). It is evident from Figure 2 that there is no correlation between the size of a TTSS batch and the task time (Figure 2a). The time spent for batches of maximum size ranges from a few seconds to the maximum sigma TTSS batch task time (~1400 seconds). On the other hand, a reasonable correlation is observed between the number of blocks (Figure 2b)/number of connections (Figure 2c) and the timings of the sigma tasks.

Knecht et al.[14] developed a computational work estimator, or weight factor, that is based on the connectivity and the size of the individual TTSS blocks. For the $i$-th TTSS batch, the weight factor is calculated as

$$a(i) = \sum_{j}^{N_{blocks}(i)} c_j(i)\, l_j(i) \qquad (10)$$

where $N_{blocks}(i)$ is the number of TTSS blocks of the $i$-th TTSS batch, $c_j(i)$ is the number of connections of the $j$-th TTSS block that belongs to the $i$-th TTSS batch, and $l_j(i)$ the number of SDs of the $j$-th TTSS block. For the CAS(20,20) test case no good correlation was found between the $a(i)$ weight factors and the sigma task timings (Figure 2d). In an attempt to better capture some of the outliers seen in the correlation graphs for the number of blocks and connections, an alternative weight factor $b(i)$ was considered, combining the connectivity and the number of TTSS blocks $N_{blocks}(i)$ in a batch. The factor $b_i$ is defined as

$$b(i) = \left( \sum_{j}^{N_{blocks}(i)} c_j(i) \right) N_{blocks}(i) \qquad (11)$$

The correlation of this weight with the sigma task timings (Figure 2e) is similar to that of the number of blocks and number of connections separately, though a few outliers now have weight factors that better reflect their computational workload.

An important step for the distribution of SDs to TTSS blocks is the organization of alpha/beta superstrings into occupation classes. This step involves a double loop over all alpha and beta occupation types (supergroups), where the full CI vector is read for each alpha/beta superstring combination. No significant time is spent for a small number of GAS spaces (i.e. small number of alpha/beta occupation types) or small CI expansions ($< 10^8$). However, significantly more time is spent when multiple GAS spaces (typically more than 6) or large active spaces are applied, and will be discussed in section 4. To reduce the amount of time spent in this generation, the double loop was decoupled and effectively parallelized by distributing the alpha occupation types over the available processors. The occupation alpha/beta connection map is built locally on each processor and is stored in a Global Array using allocated but not yet utilized memory for storing the CI and sigma vectors. The connection map is updated and in a second parallel loop the alpha/beta supergroup combinations are sorted into occupation classes.

For spin symmetric wave functions ($M_S = 0$) the innermost loops are performed in combinations rather than in SDs, as it is suggested by Knowles and Handy.[38] This approach reduces the computation effort in the inner loops by a factor of 2.

3.2 Partial AO-MO Integral Transformation

The one- and two-electron integrals and the 4-index AO-MO transformation of the two-electron integrals are performed with the optimized parallel subroutines of NWChem.[35] The two-electron integrals in MO basis are subsequently reordered to the minimal integral list that LUCIA needs, and the full set of integrals is replicated across the nodes used in the calculation. This reordering step has also been parallelized in the implementation. The choice to replicate the full set of integrals is driven by the inherent random single integral element access nature of the underlying algorithm. While each CI step may only need a subset of integrals, the full set is stored to allow for an easy 4-index transformation of the integrals needed in each MCSCF step after the molecular orbitals are rotated.

3.3 CI Eigensolver

The most time-consuming step of an MCSCF iteration is the calculation of the sigma vector (Equation 5) within the Davidson algorithm. The Davidson algorithm is an efficient procedure for the calculation of the lowest few eigenvalues by iteratively diagonalizing a subspace of a large sparse matrix. The steps that are followed are (1) the construction and diagonalization of an initial subspace matrix, which involves the calculation of an initial sigma vector, (2) calculation of the preconditioner, (3) multiplication with the inverse Hessian and diagonalization to all previous vectors, and (4) construction of the new sigma vector and diagonalization of the updated projected matrix. If convergence is not reached, return to step 2. The update of the Davidson subspace involves vector operations that are naturally parallelized along the distributed CI and sigma vectors within the Global Array framework. As stated earlier, only a subset of the CI and sigma vectors are kept in memory. The additional vectors are stored on disk using the parallel I/O (ParIO) tools of the Global Arrays Toolkit.

The update of the sigma vector is traditionally organized in either an integral-driven or a string-driven approach. LUCIA's algorithm is based on a string-driven approach. Each sigma block is obtained as a sum of alpha/alpha, beta/beta, and alpha/beta contributions. If $M_S = 0$, further simplifications are applied, the calculation of the beta/beta part is eliminated, the alpha/beta and beta/alpha contributions become equivalent, and the number of computed tasks is reduced by two. The parallel distribution of tasks proceeds as follows: each node receives a replicated copy of the one- and two-electron integrals, while the large sigma and CI vectors are stored in Global Arrays and they are distributed as described in Section 3.1. To construct its part of the sigma vector, each processor only works on the blocks of the sigma vector it has in local memory, with the coupling blocks of the CI vector fetched from the memory of remote processors using one-sided get operations and combined with the replicated integrals available locally on each node. Only non-zero sigma blocks are processed, which avoids spending computation time on redundant tasks.

3.4 Orbital Optimization Step and Outer MCSCF Iteration

The underlying vector-vector and vector-matrix multiplications of the orbital optimization and the kappa update (Eq. 9) are parallelized via the NWChem tools and take little time in the overall MCSCF calculation. As already mentioned in section 3.2, each MCSCF iteration requires a transformation of the one- and two-electron integrals to the new MO basis. Each node does this transformation locally based on a transformation matrix describing the rotation from the previous iteration MOs to the current MOs. All necessary Fock matrices needed are stored globally in a Global Array and matrix elements are updated by the processor that has them in local memory, thereby minimizing any communication.

**4. Parallel Performance**

The linear chromium trimer, $Cr_3$, was used to assess the performance of the new parallel MCSCF implementation. The Cr-Cr distance was arbitrarily set to 1.5 Å, a singlet ground state was computed, and the 6-31G* basis set was used to describe the molecular orbitals. The CI expansion was constructed using an active space of 20 electrons in 20 orbitals, CAS(20,20). An MCSCF calculation of this size is computationally not feasible for a serial code and has very large memory requirements. All $3d4s$ orbitals of the three chromium atoms were included in the active space, augmented with one occupied and one unoccupied molecular orbital to obtain the CAS(20,20) target active space. The choice of the correlated orbitals inside the active space is somehow arbitrary since these calculations were performed only for demonstrating the parallel scaling performance of the MCSCF implementation. All calculations were performed with the highest Abelian point group ($D_{2h}$), which limits the number of SDs to about 4 billion (4 267 005 808). Within this symmetry, the 20 CAS orbitals are distributed over the irreducible representations as 6 $a_g$, 1 $b_{3u}$, 1 $b_{2u}$, 2 $b_{1g}$, 5 $b_{1u}$, 2 $b_{2g}$, 2 $b_{3g}$ and 1 $a_u$. In the benchmark results the timings of a single MCSCF macroiteration with 20 Davidson CI microiterations and one extra CI iteration, used in creating the first-order density matrix, are reported. All calculations were performed using the Intel Haswell nodes on Cori supercomputer located at the National Energy Research Scientific Computing Center (NERSC). Each node has 128 GByte of memory and two 16-core Haswell processors running at 2.3 GHz, for a total of 32 cores per node. The developed version of NWChem was compiled with the Intel 16.0 compiler version, and the Global Arrays Toolkit delivering the parallel infrastructure was compiled with the MPI-PR setting. For each calculation 17 cores were used per node, 16 for the computation and one to support the MPI communication.

Table 2 and Figure 3 show the results for the CAS(20,20) case with 6 GAS spaces from 32 to 512 processors. These extra GAS spaces were created by moving all orbitals from one irrep into a different GAS space (as demonstrated in Table 1A), which resulted 888 TTSS batches available for parallel processing. The introduction of GAS spaces does not truncate the CI expansion, it only generates TTSS batches that can be efficiently distributed in different nodes, as described in Section 3.

**Table 2:** Individual times (in sec) of the MCSCF steps for one macroiteration with a CAS(20,20) for the chromium trimer. The CAS(20,20) is divided in 6 GAS spaces moving all orbitals belonging to one irreducible representation into a different GAS space (as demonstrated in Table 1A).

| Number of Processors | 32 | 64 | 128 | 256 | 512 |
|---|---|---|---|---|---|
| Time (sec) | | | | | |
| CI Generation | 56 | 80 | 43 | 23 | 18 |
| MO-AO Transformation | 24 | 18 | 12 | 8 | 6 |
| Integral Evaluation | 132 | 56 | 56 | 56 | 56 |
| CI Eigensolver (20 iterations) | 30026 | 17040 | 10331 | 5815 | 4285 |

| 1e/2e Density Matrices | 3897 | 2219 | 1442 | 1062 | 856 |
|---|---|---|---|---|---|
| Total time per MCSCF iteration | 34135 | 25104 | 15411 | 10219 | 6054 |

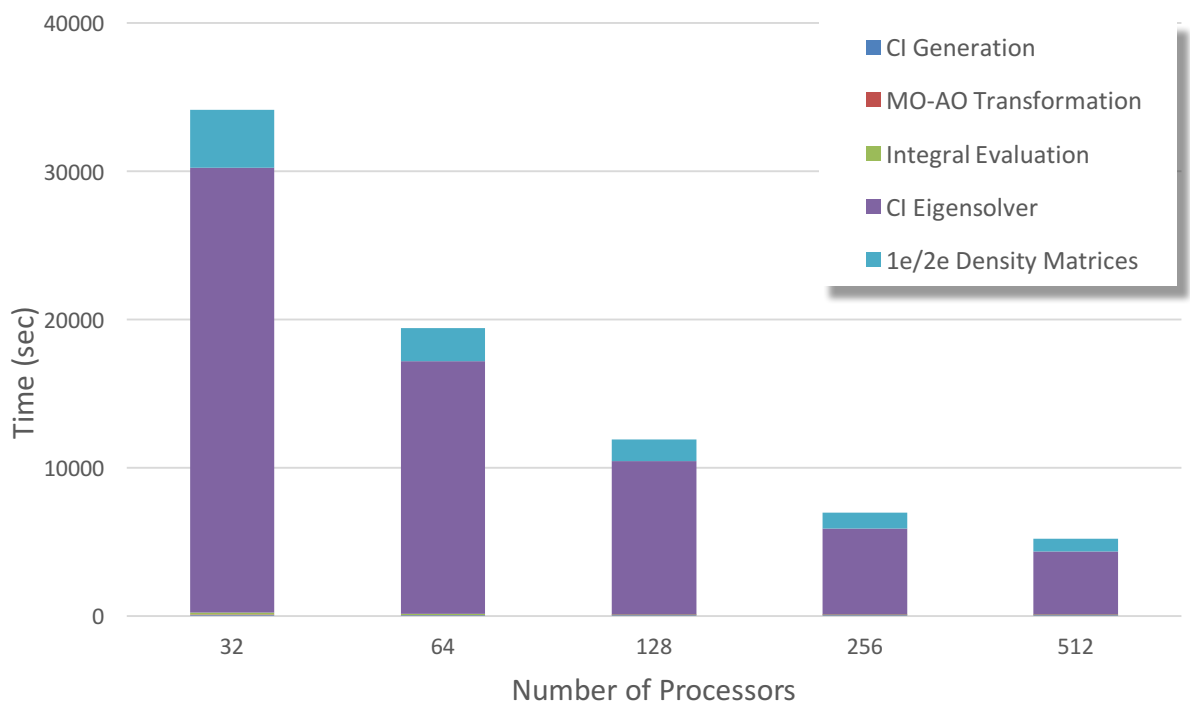

**Figure 3:** Total CPU time (in sec) and the individual contributions of one MCSCF iteration for the chromium trimer with an active space of 20 electrons in 20 orbitals and with different number of processors. The CAS(20,20) is distributed over 6 GAS spaces by moving all orbitals from one or more irreducible representations (see Table 2A as an example).

As stated before, the most time-consuming step (over 85% of the computational time) is the CI eigensolver that constructs the sigma vector in the iterative Davidson algorithm. Reasonable parallel performance was obtained scaling the CAS(20,20) calculation to 512 processors, with a speed up of 6 going from 32 to 512 processors. No perfect load-balance was achieved, which can be attributed to the small set of 888 TTSS batches or parallel tasks, each of difference size and computational complexity. To increase the number of TTSS batches that can be processes in parallel, and allow for better static load balancing, additional GAS spaces or different orbital distributions need to be introduced. In Table 3 the timing data of the CI eigensolver component for the 6 GAS spaces is compared to that of calculations using 8 GAS spaces, where each GAS space contains all orbitals of one irreducible representation. The two additional GAS spaces, each containing one orbital, generate few additional batches, and these few additional batches come at a significant computational cost. There are several factors contributing to the large increase in the computational timings when the number of GAS spaces is increased. First of all, the inner loops of the calculation of the sigma-vector are organized as loops over types of double excitations where a type specifies the four GAS spaces of the double excitation. By increasing the number of GAS spaces from 6 to 8, the number of times these loops are executed increases from 1296 to 4096. Although the number of operations executed in a given pass through the loop is reduced, the total number of operations is increased significantly. Furthermore, the use of GAS spaces with one or very few orbitals per symmetry leads to rather inefficient matrix multiplies due to matrices with one very small dimension, being equal to the number of orbitals or orbital pairs with given symmetry and GAS space.

**Table 3:** Comparison of CI eigensolver timings (in sec) in one macroiteration with a CAS(20,20) for the chromium trimer using different distributions of orbitals and GAS spaces.

| Number of Processors | 32 | 64 | 128 | 256 | 512 |
|---|---|---|---|---|---|
| 6 GAS spaces (888 batches) | 30026 | 17040 | 10331 | 5815 | 4285 |
| 8 GAS spaces (896 batches) | 42644 | 22764 | 13559 | 8736 | 4752 |

The approach of placing orbitals of each irreducible representation in a different GAS space will be able to generate at most 896 TTSS batches for the CAS(20,20) calculation considered here. As such there are not enough batches and tasks to utilize more than 896 processes. Note that, if lower symmetry would be applied the ability to generate batches for parallel processing would be limited even further. The alternative approach that was utilized is to iteratively generate new GAS spaces with only a single orbital until the number of TTSS batches significantly exceeds the number of processors available for the calculation, i.e. distributing orbitals of a single irreducible representation over multiple GAS spaces as depicted in Table 1B. The one orbital per GAS approach enabled calculations beyond 512 processors. In Table 4 the timings of calculations up to 2048 processors are presented.

**Table 4:** Individual times (in sec) of the MCSCF steps for one macroiteration with a CAS(20,20) for the chromium trimer. The number of GAS spaces is also given for all cases.

| Number of Processors | 64 | 128 | 256 | 512 | 1024 | 1024 | 2048 |
|---|---|---|---|---|---|---|---|
| Number of GAS spaces | 4 | 4 | 5 | 5 | 6 | 7 | 6 |
| Number of TTSS Batches | 288 | 288 | 1076 | 1076 | 3963 | 14502 | 3963 |
| CI Generation | 2 | 2 | 4 | 4 | 29 | 114 | 27 |
| MO-AO Transformation | 19 | 13 | 8 | 5 | 2 | 4 | 1 |
| Integral Evaluation | 4 | 4 | 8 | 8 | 15 | 34 | 14 |
| CI Eigensolver | 40967 | 26478 | 14131 | 8065 | 4873 | 6059 | 2868 |
| 1e/2e Density Matrices | 3947 | 2051 | 1844 | 941 | 803 | 872 | 555 |
| Total time per MCSCF iteration | 44939 | 28548 | 15995 | 9023 | 5722 | 7083 | 3465 |

Table 4 shows that while for smaller processor counts the one orbital per new GAS approach increases the computational cost, it does enable the overall algorithm to achieve reasonable speedup all the way to 2048 processors. Therefore, the encoded GAS distribution approach first attempts to distribute the orbitals in each irreducible representation into different GAS spaces, and only when it still does not have enough work for all available processors, it switches to distributing one orbital into a different GAS space.

## 5. Exploring CI expansions beyond CAS(20,20)

Additional test calculations were performed on larger CI expansions created by a CAS(22,22) and a CAS(24,24) to get insight into the performance of the parallel MCSCF code, and to assess the feasibility of these classes of calculations. The test calculations were run on 4096 and 8196 processors of the NERSC Cori machine. Only a single CI iteration was performed, which provides the essential information about the building of the most computationally expensive sigma vector in the new parallel CI eigensolver. To ensure maximum connectivity between sigma and the CI vector and the computational work, representative of one of the last iterations in a CI calculation without zero-valued CI blocks that would be skipped, the coefficients of all SDs in the CI expansion were given a normalized value of 1.0/(number of SDs). Given potential sparsity in the CI wave function, the calculated results are the upper bound for the computational time needed.

CASSCF calculations were performed on linear tetracene and pentacenes, which are relevant species to design organic semiconductors.[23, 42-44] For tetracene a full CI expansion that includes the complete ππ* system corresponds to a CAS(18,18). A ππ* complete active space in pentacene corresponds to 22 electrons in 22 orbitals, and a CI of 497 634 306 624 SDs. By applying symmetry restrictions ($D_{2h}$ point group), the ππ* are transformed only by four different irreducible representations, and the $^1A_g$ singlet ground state would require a CI expansion of 124 408 640 160 SDs. The number of SDs is about 30x larger than the CAS(20,20) used in the performance benchmarks in section 4. Our initial approach of moving all orbitals of a given irrep into their own GAS space did not create enough TTSS batches relative to the number of processors we intended to use for the calculation. Hence, we followed the approach discussed at the end of section 4, even though this will result in a significant computational overhead. For a calculation with 4096 processors, 7 GAS spaces were needed to generate 7788 TTSS batches, providing each processor with one or two batches to compute. Olsen and coworkers showed that the CI algorithm used here scales approximately linearly with the number of SDs,[34] which provides an excellent measure to understand possible performance degradation. The CAS(20,20) required 141 seconds per CI iteration on 2048 processors. Assuming perfect scaling, a CAS(22,22) would require 4110 seconds on 2048. The actual CAS(22,22) single CI iteration required 4212 seconds on 4096 processors, suggesting a slowdown of a factor of 2. This is attributed to significant load imbalance (the minimum and maximum time of all processors were 574 and 4212 seconds respectively). An additional eighth GAS space was needed to generate enough TTSS batches (28607) that would utilize all 8196 processors. Only a slight reduction of time was achieved by doubling the number of processors, with the CI iteration requiring 4085 seconds. This means that a full MCSCF calculation with 10 macro iterations, each with 20 CI iterations, would still require 9 days of compute time.

The chromium dimer and trimer structures are typical benchmark systems for new electronic structure theory methods. It is generally accepted that all molecular orbitals composed by the 3d and 4s atomic orbitals of the chromium atoms should be included in the active space of MCSCF calculations for $Cr_2$ and $Cr_3$. This gives rise to active spaces of CAS(12,12) and CAS(18,18) size, respectively. The chromium tetramer MCSCF with all 3d 4s atomic orbitals included would require a CAS(24,24) active space consisting of 7.3 trillion SDs (7 312 459 672 336 SDs), reduced by $D_{2h}$ symmetry and a $^1A_g$ ground state to almost one trillion SDs (914 058 513 424 SDs). Distributing the 24 orbitals in 8 GAS spaces generates 57538 TTSS batches, with each processor receiving seven or eight batches. A single CAS(24,24) CI iteration on 8196 processors required 50136 seconds. While this setup would allow the use of 20000-30000 processors, clearly, a full iterative MCSCF calculation at this scale is not yet feasible, unless the overall performance can be significantly improved.

## 6. Summary and Conclusion

With the parallel implementation presented in this work, MCSCF calculations with active spaces composed by 20 electrons in 20 orbitals can be performed routinely in small computer clusters, while calculations with almost one trillion SDs can be executed in supercomputers with thousands of processors. Such large active spaces and CI expansions are beyond the limits of current MCSCF implementations. The methodology presented in this work is based on (1) the application of the native NWChem tools for the parallelization of the MCSCF steps and (2) the generation and distribution of computational work in assembling the sigma vector of the Davidson algorithm on available processors. Fragmentation of a parent active space (CAS or GAS) into multiple GAS spaces was used to generate enough CI/sigma tasks to be distributed over processors. For active spaces as large as the CAS(20,20), good scalability up to 2048 processors has been demonstrated. For larger active spaces, such as CAS(22,22) and CAS(24,24), this approach increases significantly the number of CI tasks that have to be performed, so that smaller number of GAS spaces is preferred. It should be noted that this fragmentation approach does generate additional overhead in the calculations. Our approach shows that CI calculations with an active space composed by 22 electrons in 22 orbitals have become accessible on large parallel computing platforms, while a chromium tetramer with an active space composed by 24 electrons in 24 orbitals would require a significant increase in parallel performance.

As the seminal publication of Olsen *et al.*[34] marked the new era in multiconfigurational theory for the calculation of "exact", non-relativistic electronic energies for systems with CI expansions of one billion elements, this work sets the foundations for the "exact" treatment of larger active spaces. The new parallel implementation of MCSCF can provide benchmark results for the calibration of novel non-conventional CI methods such as DMRG or FCIQMC.

**Acknowledgements**

This work was supported by the (U.S.) Department of Energy (DOE), Office of Basic Energy Sciences, under SciDAC grant no. DE-SC0008666. This research used resources of the National Energy Research Scientific Computing Center, a DOE Office of Science User Facility supported by the Office of Science of the U.S. Department of Energy under Contract No. DE-AC02-05CH11231.

**Author Information**

Corresponding Author:

* E-mail: kvogiatz@utk.edu

* E-mail: gagliard@umn.edu

* E-mail: wadejong@lbl.gov

Current Address:

† Department of Chemistry, University of Tennessee, Knoxville, Tennessee 37996-1600, United States

¶ Max Planck Institut für Festkörperforschung, Heisenbergstraße 1, 70569 Stuttgart, Germany

References

1. Mok, D. K. W.; Neumann, R.; Handy, N. C., Dynamical and nondynamical correlation. *J. Phys. Chem.* **1996,** *100* (15), 6225–6230.
2. Handy, N. C.; Cohen, A. J., Left-right correlation energy. *Mol. Phys.* **2001,** *99* (5), 403-412.
3. Hollett, J. W.; Gill, a. P. M. W., The two faces of static correlation. *J. Chem. Phys.* **2011,** *134* (11), 114111.
4. Szalay, P. G.; Müller, T.; Gidofalvi, G.; Lischka, H.; Shepard, R., Multiconfiguration Self-Consistent Field and Multireference Configuration Interaction Methods and Applications. *Chem. Rev.* **2012,** *112* (1), 108-181.
5. Siegbahn, P. E. M., A new direct CI method for large CI expansions in a small orbital space. *Chem. Phys. Lett.* **1984,** *109* (5), 417-421.
6. Handy, N. C., Multi-root configuration interaction calculations. *Chem. Phys. Lett.* **1980,** *74* (2), 280-283.
7. Olsen, J.; Roos, B. O.; Jørgensen, P.; Jensen, H. J. A., Determinant based configuration interaction algorithms for complete and restricted configuration interaction spaces. *J. Chem. Phys.* **1988,** *89* (4), 2185.
8. Zarrabian, S.; Sarma, C. R.; Paldus, J., Vectorizable approach to molecular CI problems using determinantal basis. *Chem. Phys. Lett.* **1989,** *155* (2), 183-188.
9. Jong, W. A. d.; Bylaska, E.; Govind, N.; Janssen, C. L.; Kowalski, K.; Müller, T.; Nielsen, I. M. B.; Dam, H. J. J. v.; Veryazov, V.; Lindh, R., Utilizing high performance computing for chemistry: parallel computational chemistry. *Phys. Chem. Chem. Phys.* **2010,** *12* (26), 6896-6920.
10. Holger Dachsel; Hans Lischka; Ron Shepard; Jaroslaw Nieplocha; Harrison, R. J., A massively parallel multireference configuration interaction program: The parallel COLUMBUS program. *J. Comput. Chem.* **1997,** *18* (3), 430-448.
11. Abigail J. Dobbyn; Peter J. Knowles; Harrison, R. J., Parallel internally contracted multireference configuration interaction. *J. Comput. Chem.* **1998,** *19* (11), 1215-1228.
12. Klene, M.; Robb, M. A.; Frisch, M. J.; Celani, P., Parallel implementation of the CI-vector evaluation in full CI/CAS-SCF. *J. Chem. Phys.* **2000,** *113* (14), 5653.
13. Gan, Z.; Alexeev, Y.; Gordon, M. S.; Kendall, R. A., The parallel implementation of a full configuration interaction program. *J. Chem. Phys.* **2003,** *119* (1), 47.
14. Knecht, S.; Jensen, H. J. A.; Fleig, T., Large-scale parallel configuration interaction. I. Nonrelativistic and scalar-relativistic general active space implementation with application to (Rb–Ba)+. *J. Chem. Phys.* **2008,** *128* (1), 014108.
15. Ansaloni, R.; Bendazzoli, G. L.; Evangelisti, S.; Rossi, E., A parallel Full-CI algorithm. *Comput. Phys. Commun.* **2000,** *128* (1-2), 496-515.
16. Windus, T. L.; Schmidt, M. W.; Gordon, M. S., Parallel algorithm for integral transformations and GUGA MCSCF. *Theor. Chim. Acc.* **1994,** *89* (1), 77-88.
17. Müller, T., Large-Scale Parallel Uncontracted Multireference-Averaged Quadratic Coupled Cluster: The Ground State of the Chromium Dimer Revisited. *J. Phys. Chem. A* **2009,** *113* (45), 12729-12740.
18. White, S. R., Density matrix formulation for quantum renormalization groups. *Phys. Rev. Lett.* **1992,** *69* (19), 2863-2866.
19. Schollwöck, U., The density-matrix renormalization group. *Rev. Mod. Phys.* **2005,** *77*, 259-315.
20. Chan, G. K.-L.; Sharma, S., The Density Matrix Renormalization Group in Quantum Chemistry. *Annu. Rev. Phys. Chem.* **2011,** *62*, 465-481.

21. Zgid, D.; Nooijen, M., The density matrix renormalization group self-consistent field method: Orbital optimization with the density matrix renormalization group method in the active space. *J. Chem. Phys.* **2008,** *128* (14), 144116.
22. Ghosh, D.; Hachmann, J.; Yanai, T.; Chan, G. K.-L., Orbital optimization in the density matrix renormalization group, with applications to polyenes and β-carotene. *J. Chem. Phys.* **2008,** *128* (14), 144117.
23. Gidofalvi, G.; Mazziotti, D. A., Active-space two-electron reduced-density-matrix method: Complete active-space calculations without diagonalization of the N-electron Hamiltonian. *J. Chem. Phys.* **2008,** *129* (13), 134108.
24. Poelmans, W.; Van Raemdonck, M.; Verstichel, B.; De Baerdemacker, S.; Torre, A.; Lain, L.; Massaccesi, G. E.; Alcoba, D. R.; Bultinck, P.; Van Neck, D., Variational Optimization of the Second-Order Density Matrix Corresponding to a Seniority-Zero Configuration Interaction Wave Function. *J. Chem. Theory Comput.* **2015,** *11* (9), 4064-4076.
25. Fosso-Tande, J.; Nguyen, T.-S.; Gidofalvi, G.; DePrince, A. E., III, Large-Scale Variational Two-Electron Reduced-Density-Matrix-Driven Complete Active Space Self-Consistent Field Methods. *J. Chem. Theory Comput.* **2016,** *12* (5), 2260-2271.
26. Booth, G. H.; Thom, A. J. W.; Alavi, A., Fermion Monte Carlo without fixed nodes: A game of life, death, and annihilation in Slater determinant space. *J. Chem. Phys.* **2009,** *131* (5), 054106.
27. Blunt, N. S.; Smart, S. D.; Kersten, J. A.-F.; Spencer, J. S.; Booth, G. H.; Alavi, A., Semi-stochastic full configuration interaction quantum Monte Carlo: Developments and application. *J. Chem. Phys.* **2016,** *142* (18), 184107.
28. Thomas, R. E.; Sun, Q.; Alavi, A.; Booth, G. H., Stochastic Multiconfigurational Self-Consistent Field Theory. *J. Chem. Theory Comput.* **2015,** *11* (11), 5316-5325.
29. Li Manni, G.; Smart, S. D.; Alavi, A., Combining the Complete Active Space Self-Consistent Field Method and the Full Configuration Interaction Quantum Monte Carlo within a Super-CI Framework, with Application to Challenging Metal-Porphyrins. *J. Chem. Theory Comput.* **2016,** *12* (3), 1245-1258.
30. Ma, D.; Li Manni, G.; Gagliardi, L., The generalized active space concept in multiconfigurational self-consistent field methods. *J. Chem. Phys.* **2011,** *135* (4), 044128.
31. Ivanic, J., Direct configuration interaction and multiconfigurational self-consistent-field method for multiple active spaces with variable occupations. I. Method. *J. Chem. Phys.* **2003,** *119* (18), 9364.
32. Vogiatzis, K. D.; Li Manni, G.; Stoneburner, S. J.; Ma, D.; Gagliardi, L., Systematic Expansion of Active Spaces beyond the CASSCF Limit: A GASSCF/SplitGAS Benchmark Study. *J. Chem. Theory Comput.* **2015,** *11* (7), 3010-3021.
33. Knecht, S.; Jensen, H. J. A.; Fleig, T., Large-scale parallel configuration interaction. II. Two- and four-component double-group general active space implementation with application to BiH. *J. Phys. Chem.* **2010,** *132* (1), 014108.
34. Olsen, J.; Jørgensen, P.; Simons, J., Passing the one-billion limit in full configuration-interaction (FCI) calculations. *Chem. Phys. Lett.* **1990,** *169* (6), 463-472.
35. Valiev, M.; Bylaska, E. J.; Govind, N.; Kowalski, K.; Straatsma, T. P.; Dam, H. J. J. V.; Wang, D.; Nieplocha, J.; Apra, E.; Windus, T. L.; Jong, W. A. d., NWChem: A comprehensive and scalable open-source solution for large scale molecular simulations. *Comput. Phys. Commun.* **2010,** *181* (9), 1477-1489.
36. F. Aquilante, J. A., R. K. Carlson, L. F. Chibotaru, M. G. Delcey, L. De Vico, I. Fdez. Galvan, N. Ferre, L. M. Frutos, L. Gagliardi, M. Garavelli, A. Giussani, C. E. Hoyer, G. Li Manni, H. Lischka, D. Ma, P. Malmqvist, T. Muller, A. Nenov, M. Olivucci, T. B. Pedersen, D. Peng, F. Plasser, B. Pritchard, M. Reiher, I. Rivalta, I. Schapiro, J. Segarra-Marti, M. Stenrup, D. G. Truhlar, L. Ungur, A. Valentini, S. Vancoillie, V. Veryazov, V. P. Vysotskiy, O. Weingart, F. Zapata, and R. Lindh, MOLCAS 8: New Capabilities for Multiconfigurational Quantum Chemical Calculations across the Periodic Table,. *J. Comput. Chem.* **2016,** *37* (5), 506-541.

37. Roos, B., A new method for large-scale CI calculations. *Chem. Phys. Lett.* **1972,** *15* (2), 153-159.
38. Knowles, P. J.; Handy, N. C., A new determinant-based full configuration interaction method. *Chem. Phys. Lett.* **1984,** *111* (4-5), 315-321.
39. Davidson, E. R., The iterative calculation of a few of the lowest eigenvalues and corresponding eigenvectors of large real-symmetric matrices. *J. Comput. Phys.* **1975,** *17* (1), 87-94.
40. Roos, B. O.; Siegbahn, P. E. M., In *Methods of Electronic Structure Theory*, Schaefer III, H. F., Ed. Plenum Press: New York, 1977; Vol. 3, p 277.
41. Nieplocha, J.; Palmer, B.; Tipparaju, V.; Krishnan, M.; Trease, H.; Apra, E., Advances, Applications and Performance of the Global Arrays Shared Memory Programming Toolkit. *International Journal of High Performance Computing Applications* **2006,** *20* (2), 203-231.
42. Hachmann, J.; Cardoen, W.; Chan, G. K.-L., Multireference correlation in long molecules with the quadratic scaling density matrix renormalization group. *J. Chem. Phys.* **2006,** *125* (14), 144101.
43. Horn, S.; Plasser, F.; Müller, T.; Libisch, F.; Burgdörfer, J.; Lischka, H., A comparison of singlet and triplet states for one- and two-dimensional graphene nanoribbons using multireference theory. *Theor. Chem. Acc.* **2014,** *133* (8), 1511.
44. Lee, J.; Small, D. W.; Epifanovsky, E.; Head-Gordon, M., Coupled-Cluster Valence-Bond Singles and Doubles for Strongly Correlated Systems: Block-Tensor Based Implementation and Application to Oligoacenes. *J. Chem. Theory Comput.* **2017,** *13* (2), 602–615.